# Congestion Control in High-speed Networks Using the Probabilistic Estimation Approach

Shahram Jamali[1]  . Mir Mahmoud Talebi[2] . Reza Fotohi[3]

[1]Computer Engineering Department, University of Mohaghegh Ardabili, Ardabil, Iran

[2]Department of Computer Engineering Germi Branch, Islamic Azad University Germi, Iran

[3]Faculty of Computer Science and Engineering, Shahid Beheshti University, Tehran, 1983969411, Iran

**Correspondence**
Faculty of Computer Science and Engineering, Shahid Beheshti University, Tehran, 1983969411, Iran

Email: Fotohi.reza@gmail.com;
R_fotohi@sbu.ac.ir

**Abstract**
Nowadays, the bulk of Internet traffic uses TCP protocol for reliable transmission. But the standard TCP's performance is very poor in High Speed Networks (HSN) and hence the core gigabytes links are usually underutilization. This problem has roots in conservative nature of TCP, especially in its Additive Increase Multiplicative Decrease (AIMD) phase. In other words, since TCP can't figure out precisely the congestion status of the network, it follows a conservative strategy to keep the network from overwhelming. We believe that precisely congestion estimation in the network can solve this problem by avoiding unnecessary conservation. To this end, this paper proposes an algorithm which considers packet loss and delay information jointly and employs a probabilistic approach to accurately estimation of congestion status in the network. To examine the proposed scheme performance, extensive simulations have been performed in the NS-2 environment. Simulation results reveal that the proposed algorithm has better performance than existing algorithms in terms of bottleneck utilization, stability, throughput and fairness.

## 1 | INTRODUCTION

TCP (Transmission Control Protocol) is normally utilized in current networks providing reliable end-to-end data communication. Evolving the web to contain several long distances and high-speed network routes challenged the TCP protocol behaviour. Massive bandwidth and delay product (BDP) are characteristics of these networks that provide the total number of packets required in flight though maintaining the band width absolutely utilised [1-5]. In standard TCP like TCP-New Reno TCP-Reno, and TCP- SACK, TCP extends its window one per roundtrip time (RTT). Although the standard TCP was considerably effective to perform congestion avoidance and control in preventing severe congestion in the low-speed networks, the standard TCP is not suitable for networks with highspeed where the additional increment multiplicative decrement

(AIMD) algorithm is too cautious for obtaining full bandwidth use rapidly while is too extreme for recovering from per packet loss case. To overcome the low-performance problem, the standard TCP's AIMD algorithm needs to be improved for high-speed networks. Many high-speed end-to-end congestion management Schemes have been provided, so far. These approaches are classified into three groups of delay-based, loss-based, and their hybrid (combine loss-delay-based).

Within loss-based protocol, an Additive Increase Multiplicative Decrease (AIMD) mechanism of TCP congestion prevention phase is modified to rapidly increment and gradually reduce the congestion window compared to TCP-Reno, to obtain high throughput in high-speed networks adjusting the size of congestion window through resulting in intentional packet losses. The examples include Scalable TCP, High-speed TCP (HSTCP, for short), BIC, early version of TCP-Westwood (TCPW), CUBIC. RTT is used in delay-based outlines for predicting network congestion prior to packet losses indicating superior behaviour in fairness and efficiency.

FAST-TCP [6] and TCP-Vegas [7] are examples in this regard. Nevertheless, it is indicated that they have strictly corrupted behaviours when competing with loss-based TCP protocols. Hybrid approaches combining loss-based and delay-based properties were recently proposed including TCP-Illinois, Hybrid Congestion Control TCP (HCC TCP) [8], and Compound TCP (CTCP) [9]. Based on the RTT-estimated congestion level measurement, they adaptively change their congestion control modes (delay-based and loss-based). Although these approaches achieve delay-based schemes' high throughput efficiency in case of non-congested network, and offer friendliness to loss-based schemes in case of congested network, there some deficiencies in most of them in different aspects including responsiveness, TCP-friendly, and fairness. Since there are no proposed approaches mainly better than the other approaches and other effective methods should be studied, it is still required to develop novel high-speed TCP variants.

Here, we propose yet another variant of TCP, called probability hybrid congestion control TCP (PHCC TCP), for networks with high speed that relies on predictions based on probabilities. In other words, we use the conditional probability and Bayes theory [10] is utilized for estimating the congestion window. Furthermore, unlike other proposed hybrid methods that, two indicators used separately delay and packet loss. The protocol utilizes the delay and loss information as the simultaneously congestion indicator to cooperatively set the window size in order to meet the design needs on efficiency, TCP-friendliness, fairness, and overtakes the loss/delay-based TCP and hybrid TCP protocols in high-speed networks.

The rest of this study is adjusted as: Section 2 explains the relevant studies. In Section 3, we explain the proposed method the PHCC TCP algorithm. Section 4 includes evaluations of simulation experiment results. Ultimately, we provide the conclusion and talk about the future work in Section 5.

## 2 | Related works

This section lists the methods that have addressed the issue of congestion control so far. Each method is briefly reviewed.

In [11], the hybrid optimization algorithm for sensor networks is used in a closed experimental experiment based on a rapid congestion control scheme. First, using the waiting delay, received signal strength and mobility as multiple inputs of this algorithm, we select the best intermediate node in the optimization algorithm, which increases the life of the network to reduce congestion. Then, using the modified gravitational search algorithm, we calculate the path for the node to sink, which provides better routing. The simulation results show that the proposed method can control congestion by reducing energy consumption, missing packets and number of steps and increasing network life compared to the current methods.

Increasing the throughput and reducing the latency of very fast and variable network connections is targeted in the method proposed in this paper and we classify them based on the way they control congestion. This article also discusses the application of algorithms. In addition, this paper discusses future research that could help develop and deploy better-performing algorithms. They also offered future research pathways, such as dealing with a higher degree of diversity, the interaction of CCAs in a common bottleneck, and ways for synchronous research, such as CCAs defined by software networks and virtualization of network performance [12].

A hybrid congestion control-based approach that uses latency information as the main congestion indicator and loss information as the second congestion indicator to jointly adjust window size to meet design needs for TCP justice and performance. Better than other standards such as TCP and other types of TCP in high speed networks. Due to the synergy of delay-based strategy and loss-based strategy, HCC TCP is a hybrid congestion control scheme. The proposed mechanism is very practical for high-speed networks and has performed better than the methods compared to them in the paper [13].

Inspired by traction-passing algorithms in other fields, we proposed a methodology for active congestion control in this paper, which explicitly calculates rates independently of congestion signals. For example, modern low-density parity decoding algorithms improve the convergence time by a factor of 7 compared to explicit reactive speed control protocols such as RCP. This rapid convergence significantly reduces the tail end current in high-speed networks. We also show that in such cases, the active algorithms obtain far fewer tail FCTs due to their rapid convergence than the reaction algorithms. For example, realistic load simulations in a 100 Gbps network show that PERC achieves better efficiency than RCP [14].

For deployment in high-speed and long-distance networks as well as conventional networks, we have introduced a new AIMD congestion control algorithm, H-TCP, which works much better than other existing methods. H-TCP extends the AIMD strategy to significantly improve response and scale product performance with latency in bandwidth and queue supply level in the network. Also in this paper, we consider the problem of congestion control protocol design for deployment in high-speed and long-distance networks, which results in better performance of the proposed protocol through simulation measurements in a wide range of network conditions [15].

A hybrid method using bottleneck bandwidth and distance propagation time (RCP-BBR) is proposed as an alternative to UDP for congestion control. The proposed method achieved efficient control of congestion, low latency, high throughput and efficient contact personalization ratio with efficient use of bandwidth as bottleneck bandwidth and back-and-forth release time. According to the simulation results, the proposed protocol achieves better throughput through UDP in stable networks. In addition, in unstable and remote networks, the introduced method achieved smaller queues in deep buffers and less delays compared to UDP, which performs poorly by keeping the delays above the contact drop threshold [16].

End-to-end algorithms propose an attractive approach to Internet congestion control, both in simplicity and scalability. In this section, we present some of the most relevant work on TCP end-to-end congestion control. Recall that the only signals of network congestion available to an end-to-end algorithm are packet losses and latency variations. Therefore, research has been focused on three types of algorithms [17]:

- Loss-based algorithms rely on packet losses alone to react to network congestion;
- Delay-based algorithms use delay measurements alone to infer router queue occupancy and act before heavy congestion occurs;
- Hybrid-based algorithms use techniques from both loss-based and delay-based algorithms; their rationale is that with more information, an end-to-end algorithm can infer the state of the network more accurately and make better decisions.

Kelly proposed Scalable TCP (STCP). Scalable TCP modifies the congestion control algorithm. The congestion window is reduced by this algorithm for each packet loss by a factor of 1/8 instead of Standard TCP's 1/2, till stopping the packet loss. This reduces the recovery time on 10 Gbit/s links from 4 h and more (utilizing Standard TCP) to less than 15 s for the round-trip time of 200 ms. High-speed TCP (HSTCP) uses an adjusted AIMD by a convex function of the current congestion window size in case the multiplicative decrease factor and linear increase factor are modified. HSTCP has a behaviour similar to standard TCP, in case of the congestion window less than some cut-off value. This window size-based TCP compatibility is supported by most of high-speed TCP variants. Several of these protocols in competition with shares of the bottleneck link possess various RTT delays, use of bandwidth efficiency cannot be fair [18].

A binary search algorithm is used by BIC TCP for growing window to the mid-point within the last window size (namely max). Here, TCP contains the last window size (namely min) and a packet loss and there is no loss over a RTT period. BIC TCP adjusts the mid-point as the new min and accomplishes another "binary-search" possessing the max and min windows. It affects the window growth really fast if the current window size is far from the accessible capacity of the path. Hence, its window increase is gradually reduced when it is close to the existing capacity (the former loss). This concave function makes BIC TCP very stable and simultaneously highly scalable. Although the performance of BIC TCP is better than old presented protocols. However, also the problem of RTT unfairness in this protocol remains. Afterwards, an improved form of BIC TCP called CUBIC TCP is established to enhance the RTT-fairness of its performance. In fact, CUBIC TCP is a kind of more structured and less aggressive BIC TCP where the window size is determined as a cubic function of time from the last congestion event. HTCP like CUBIC TCP for calculation of the current congestion window size utilizes the elapsed time ($\Delta$) from the final congestion. The HTCP's window growth function is a quadratic function of $\Delta$ where HTCP is unique and sets the reduction fact or by a function of RTTs planned for estimating the queue size in the current flow's network path. The increment of window size for HTCP, HSTCP and STCP protocols, is still fast even the close network to the congestion. The more flows competition and reduced throughput are obtained by the congestion in network [19].

TCP Vegas determines the difference ($\delta$) between the actual throughput and estimated throughput in terms of round-trip delays. In case, $\delta$ is less than a low threshold $\alpha$, TCP Vegas trusts the route as the non-congested, therefore, it increments the sending rate. In case of $\delta$ higher than an upper threshold $\beta$ as a robust evidence of congestion, TCP Vegas decreases the window size of sending. Then, TCP Vegas keeps the current directing window size. The estimated throughput is determined by division of the current congestion window by the least RTT normally containing the delay, till finding the congested path. TCP Vegas calculates the actual throughput for each RTT by division of the number of packets directed by the tested RTT. FAST TCP is a Vegas's high-speed descendant. Even though FAST TCP builds upon the principles of Vegas, it increments the congestion window more violently to achieve good efficiency in high-speed networks. This protocol keeps queue occupying at paths for a small value to direct the network around full bandwidth use and obtain a greater average throughput. Furthermore, FAST TCP can converge to the equilibrium mode rapidly while not suffering the RTT unfairness problem. FAST TCP splits the congestion window and enters loss recovery similar to TCP for packet losses. However, it also includes some weaknesses. Given that FAST TCP is a delay-based method, it utilizes the RTTs for congestion and its throughput behaviour is considerably influenced by the traffic of reverse path, and its throughput decreases as the queuing delay increases on the reverse path [20].

TCP-Illinois follows an AIMD algorithm, but uses delay estimates to set the increase and decrease congestion window parameters. If TCP-Illinois does not detect queuing delay (i.e., network congestion), the increase parameter is set to the maximum value, making the congestion window grow quickly. As the queuing delay starts to build up, the increase parameter is then gradually decreased, making the congestion window grow more slowly. It utilizes the loss information to determine the direction of window change and the delay information is used for adjusting the pace of window size change. Compound TCP (CTCP) is planned to violently set the congestion window of the sender to enhance TCP for connections with huge bandwidth-delay products while not harming fairness (since it is able to occur with HSTCP). CTCP regards two congestion windows: a delay-based window and a normal AIMD window. The sum of these two windows is used to calculate the actual sliding window size. Thus, these methods use the benefits of both the delay-based and loss-based approaches. Nevertheless, since RTTs are used in the delay-based constituents to measure the congestion, similar to FAST TCP, reverse path traffic also affects their throughput behaviour [21].

This method is similar to previous methods have shortcomings that they are not paying attention to the nature of packet loss and of its severity. The main emphasis of this approach is based on the delay, and packet loss, only to switch to a strict mode is used to reduce the size of the window. This algorithm will decrease throughput when the packet loss probability increases. It also suffers the TCP-friendliness problem.

## 3 | The PHCC TCP protocol

Congestion control in High-Speed Networks (HSN) is designed in the following section using the probabilistic estimation approach (PEA). The PHCC TCP protocol consists of five phases, such as the overview of the PHCC TCP protocol in the HSN is discussed in Sect. 4.1. Architecture and Congestion probability model is discussed in Sect. 4.2, Delay and loss probability functions is discussed in Sect. 4.3; Joint congestion window estimation model is discussed in Sect. 4.4. and Congestion window algorithm is discussed in Sect. 4.5.

### 3.1 Phase 1: Overview of the PHCC TCP protocol

The combination methods, with the synergy of delay-based method and loss-based method, can solve many shortcomings of both the loss-based and delay-based approaches. Therefore, PHCC TCP also adopted the method that uses the synergy of the loss-based and delay-based approach as jointly to realize the congestion control for high-speed networks. However recent studies show the terms of use of delay and loss information has a considerable impact on throughput, fairness, sensitivity to buffer size, TCP friendliness, and etc. Hence, in the use of indicators and define the functions must be performed carefully to any factor that greatly affects the network traffic should not be forgotten. Since the discovery of all the factors involved in congestion, the amount and how it is virtually impossible, in this paper, instead of using deterministic functions of mathematical, probability theory is used to estimate the traffic condition. This mechanism fundamentally differentiates PHCC TCP from others.

### 3.2 Phase 2: Architecture and Congestion probability model

As shown in Fig. 1, the delay could be two main reasons, congestion control policy applied and other problems (such as communication link problems and traffic of other flows). Packet loss in addition to other problems (connection failure, etc.) could be due to the inappropriate congestion control policy. PHCC in the first step, calculate delay-based probability and loss-based probability from delay and loss information, and in the second step, the window control strategy calculate from the estimates of the delay and loss probability is realized by the joint congestion control component.

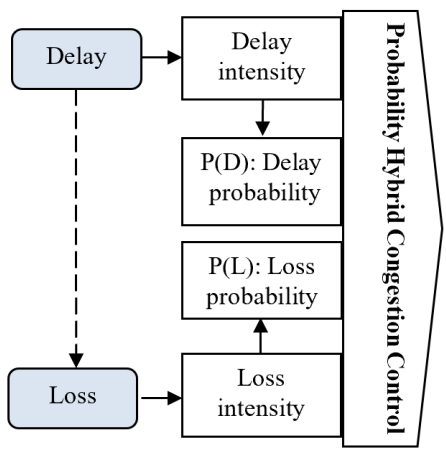

**FIGURE 1** PHCC TCP architecture.

According to Fig. 1 and using probability theory the probability space of network congestion in Fig. 2 for a flow could be drawn. In Fig. 2, the congestion probability space is shown with M, that can be the value of "zero" (the absence of any congestion), and the "One" (Timeout or loss) or a value between "zero" and "one" according to the hardness of congestion.

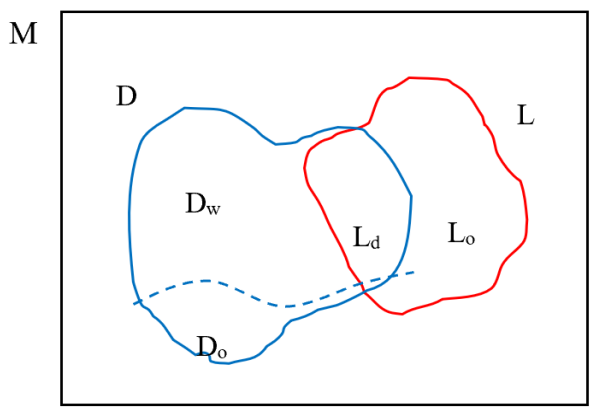

**FIGURE 2** Probability space of network congestion.

From fig. 2, Delay D consists of delays caused by congestion window changes $D_w$ and delays caused by other factors $D_o$. Similarly, Loss L consists of losses caused by delay $L_d$ and losses caused by other factors $L_o$.

### 3.3 Phase 3: Delay and loss probability functions

According to the mechanism shown in Figure 1, the window size is a function of the complement probability of delay $1 - P(D)$ and packet loss complement probability $1 - P(L)$. Window size can be written as:

$$WindowSize = \left( f\left(1 - P(D), 1 - P(L)\right) \right) \tag{1}$$

As illustrated in Fig. 3, using a Bayesian network, we calculate the estimation of delay probability $P(D^{est})$ as follows:

$$P(D^{est}) = P(D^{old}).P(D|D^{old}) + (1 - P(D^{old})).P(D|\sim D^{old}) \tag{2}$$

The old delay probability $P(D^{old})$ is the mean value of the delay probability is obtained from the previous RTT. $P(D|D^{old})$, the conditional probability or likelihood, is the degree of belief in delay D, given that the proposition delay of old $(D^{old})$ is true. Likewise, $\sim D^{old}$ means that the proposition "old delay" is false. Similarly, the estimation of packet loss probability $P(L^{est})$ can be written as

$$P(L^{est}) = P(L^{old}).P(L|L^{old}) + (1 - P(L^{old})).P(L|\sim L^{old}) + P(D^{old}).P(L|D^{old}) + (1 - P(D^{old})).P(L|\sim D^{old}) \tag{3}$$

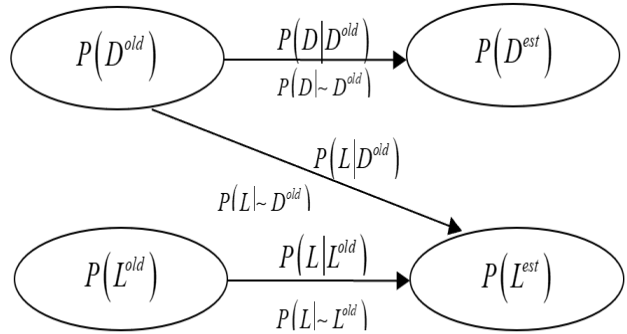

**FIGURE 3** Bayesian model of delay and loss.

### 3.4 Phase 4: Joint congestion window estimation model

Based on the previous sections describes, the estimation of the congestion window according to Fig. 3 are calculated. In fig. 3 the delay could be two different reasons, inappropriate increase the size of the current window and affects the other flows traffic. Packet loss can be due to the delay and other factors. In order to overlap of the two indicators delay and packet loss, to determine size of the congestion window $W_i^{est}$ for source of $i$, Firstly, the appropriate estimation of window size $W(D_i^{est}), W(L_i^{est})$ based on the theory Bayes for both delay and packet loss will be calculated separately. The estimation of delay Probability $W\left(D_i^{est}\right)$ and estimation of packet loss probability $W\left(L_i^{est}\right)$ can be also calculated from Eq. (2), (3).

$$W_i^{est}(D) = \left(1 - P\left(D_i^{est}\right)\right).W\left(D_i^{tar}\right) \quad (4)$$

$$W_i^{est}(L) = \left(1 - P\left(L_i^{est}\right)\right).W\left(L_i^{tar}\right) \quad (5)$$

Without the effect from the queuing delay on the reverse path, the source $i$ can achieve full utilization of available bandwidth on the forward path, let be the current delay, the queuing delay of source $i$ is $D_{queue}$. Each source computes its delay-based target window periodically according to:

$$W\left(D_i^{tar}\right) = \left(D_i * \frac{\alpha_i}{D_{queue}}\right) \quad (6)$$

Size of the target window for loss-based strategy $W\left(D_i^{tar}\right)$ in fact, the window size is a reference to packet loss. In other words, if a packet loss is detected during the growth of the window size, the source updates the loss-reference to the current window size. Finally, the congestion window size is estimated as

$$W_i^{est}(D,L) = \left(\frac{W_i^{est}(D) + W_i^{est}(L)}{2}\right) \quad (7)$$

### 3.5 Phase 5: Congestion window algorithm
In this subsection, we show how the delay-based strategy and the loss-based strategy are used simultaneously in the joint control phase of PHCC TCP. At start-up, PHCC TCP relies on the delay-based algorithm to increase the window size [22-25]. Firstly, we set a delay threshold minimum value $D_{min}^{tr}$ to estimate the congestion by use of slow start threshold *ssthresh* and expected bandwidth value $bw_{expect}$ from of TCP Vegas that can be written as

$$D_{min}^{tr} = \left(\frac{ssthresh}{bw_{expect}}\right) \quad (8)$$

Figure 4 shows three phase of congestion state, in first phase, if $Delay < D_{min}^{tr}$, it indicates that the queuing is light and available bandwidth is not used fully, for rapidly increase the window size, the additive multiplicative increase scheme can be used. We set a delay threshold maximum value $D_{max}^{tr} = \{D_{delay}^{tr}, D_{loss}^{tr}\}$ for divide the remained state ($Delay > D_{min}^{tr}$) in two phase; congestion control phase: if $D_{min}^{tr} < Delay < D_{max}^{tr}$ and critical phase: when $D_{max}^{tr} < Delay < Timeout$. Delay reference $D_{loss}^{tr}$ is updated for every loss detect. The delay threshold $D_{delay}^{tr}$ is calculated by using the values of sequence number *seqno*, last ack number *lastack* and buffer size *BFS* that is obtained from expected bandwidth $bw_{expect}$ and actual bandwidth $bw_{actual}$ functions of TCP Vegas as shown below.

$$BFS = \left(\left(bw_{expect} - bw_{actual}\right) * D_{min}\right) \tag{9}$$

$$D_{delay}^{tr} = \left(\left(seqno - lastack\right) * \left(\frac{Delay - D_{min}}{BFS}\right)\right) \tag{10}$$

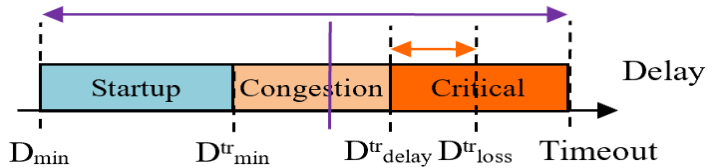

**FIGURE 4** Three phase of congestion control state.

Congestion phase in Fig. 4 is main phase of our algorithm. In this phase we calculate the estimation window for congestion control by Eq. 7 per RTT. If any loss detects or timeout event in each of three phases. The algorithm decreases the window size in half and check policy parameter for select correct phase. Figure 5 shows the PHCC TCP algorithm flowchart.

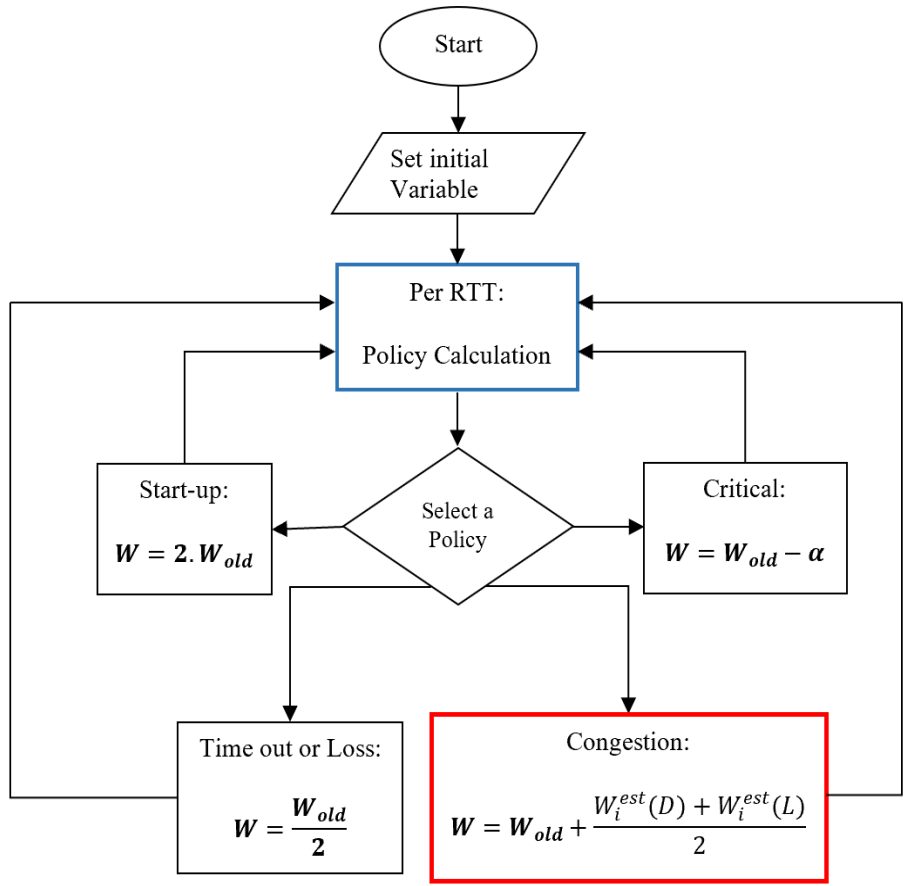

**FIGURE 5** PHCC TCP algorithm flowchart.

## 4 | Evaluating the Performance

This section evaluates the performance of PHCC TCP using the NS 2.35 simulator [26-29]. The TCP Vegas agent in NS-2 are modified to implement PHCC TCP. We implemented wide simulation experiments to assess the PHCC TCP and made a comparison between its performance with other old protocols such as: TCP Vegas, TCP Reno, STCP, HSTCP, HTCP, TCP-Illinois, BIC-TCP, FAST TCP and HCC TCP, we used a dumb-bell network model with a single bottleneck shared link as shown in Fig. 6. The sources, routers and destinations links each with 200Mbs bandwidth and 20ms delay. Router queue discipline is first-in-first-out (FIFO) selected. In all tests, the size of data packets is 1000 bytes. For other TCP protocol, their default parameter setting is used.

**PHCC Behavior as a single traffic flow:** As an initial evaluation, performance of the proposed algorithm for different phases (slow start, congestion avoidance and achieve maximum throughput) tested. Average throughput of the proposed method is presented in Fig 6. It can be seen that when the RTT is 2ms, as shown in Fig. 6(a), the algorithm achieves the optimal and maximum use of available bandwidth and the average throughput increase faster. As shown in Fig. 6(b), when the RTT is 20ms, the average throughput of PHCC TCP Protocol grow slowly by controlling the congestion in the network.

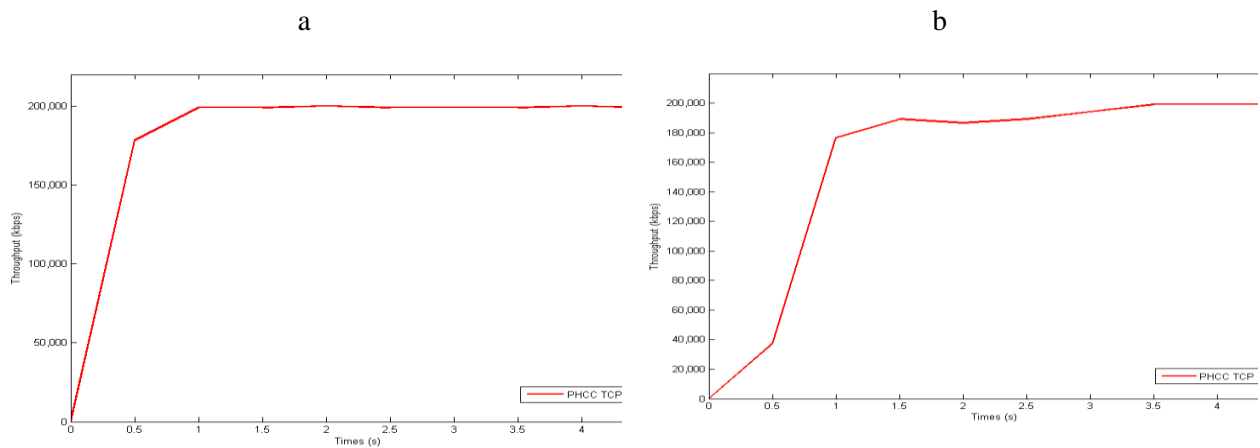

**FIGURE 6** Throughput of PHCC TCP algorithm as a single traffic, (a) RTT(2ms) and (b) RTT (20ms).

As shown in Fig. 7, the algorithm has a behavior such as other protocols with a difference threshold in slow start phase, in adapting phase TCP PHCC to achieve an optimum situation of two major indicator used, In both indicator, the probability theory used to adjust the window. This algorithm in stability phase is trying to avoid the congestion and use of the maximum available bandwidth. With this description, the PHCC TCP in contrast to other presented methods, in use of delay index, except of delay size, the expected based on Bayes' theorem is also used.

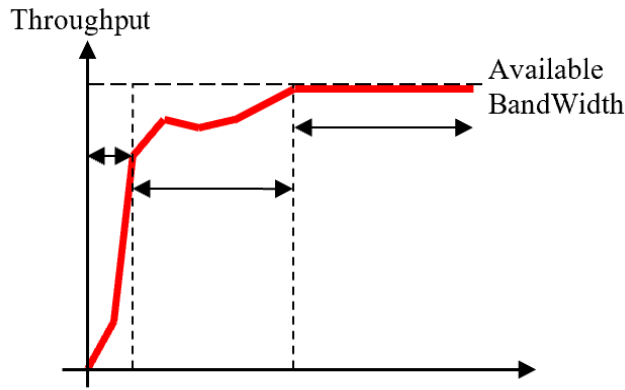

**FIGURE 7** Three phase of PHCC throughput grow.

**Evaluation of throughput:** One of the indicators that proposed algorithm is used to congestion control, is packet loss, in Fig. 8, the range of throughput to compete with the other two protocol (a normal TCP flow and a fast TCP flow) is shown. According to this algorithm, the packet loss due to congestion is not considered a hundred percent, has a Better performance than other algorithms and if recognize that although packet loss, but delay time is low and traffic flow is normal, applying strict policy is avoided. The above experiments on two scenarios assuming packet loss probability of 0.01 and 0.00001, we repeat the results of which are shown in Fig. 8. The result shows that the rival algorithms even when packets are lost is rarely, because packet loss are considered as a binary indicator, heavily reduces the congestion window size and causing severe fluctuations in throughput.

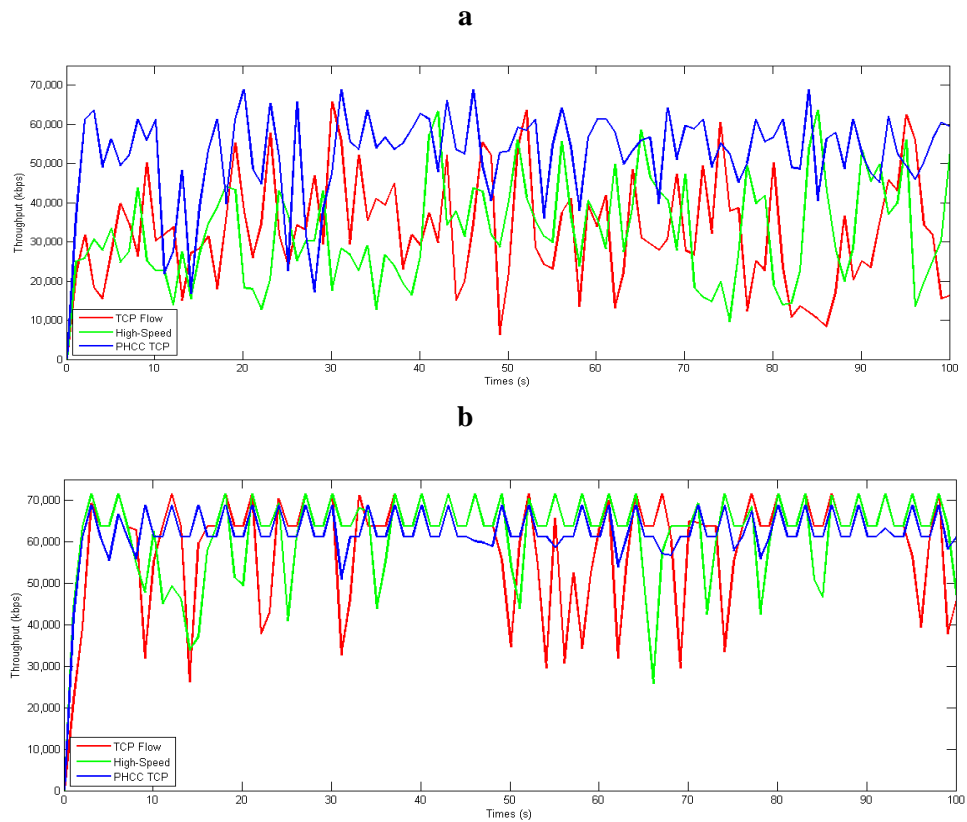

**FIGURE 8** Rate dynamics of PHCC TCP throughput with loss probability (a) loss probability is 0.01 and "(b) loss probability is 0.00001.

To evaluate the throughput and efficiency of the proposed method is to change the buffer size, the simulation run with different buffer values, simulation time was 300 seconds. Figure 9 represents the average throughput for different buffer sizes of 100-4000 packets. We can see an increase in the average throughput of all protocol with buffer size growth, protocols that are planned for high-speed networks, uniformly better throughput than protocols like TCP Vegas and TCP Reno that are not planned for high-speed networks. It is observed that throughput of TCP FAST with the buffer size of less than 400 packages, falls rapidly. The algorithm, except when the buffer size is less than 300, in other cases, as well as the available buffer used and adapted with it. PHCC TCP acts well on throughput behaviour compared to other high-speed TCP variants nearly in all buffer size cases.

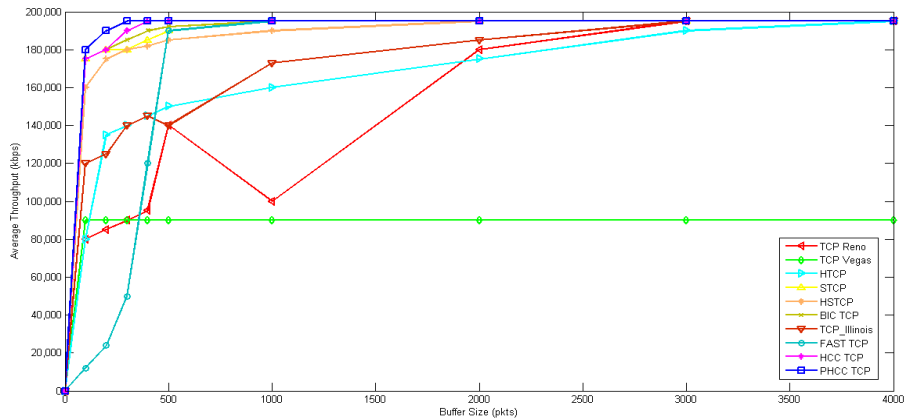

**FIGURE 9** Comparison of the buffer size vs throughput.

In further experiments, the throughput of the proposed algorithm with the normal algorithms and algorithms are presented for high-speed network compared to the probability of packet loss. The results of which are shown in Fig. 9, although the proposed algorithm has not the best performance compared to other algorithms, but this algorithm in total in acceptable condition. FAST TCP algorithm is better in this test due to its delay is based.

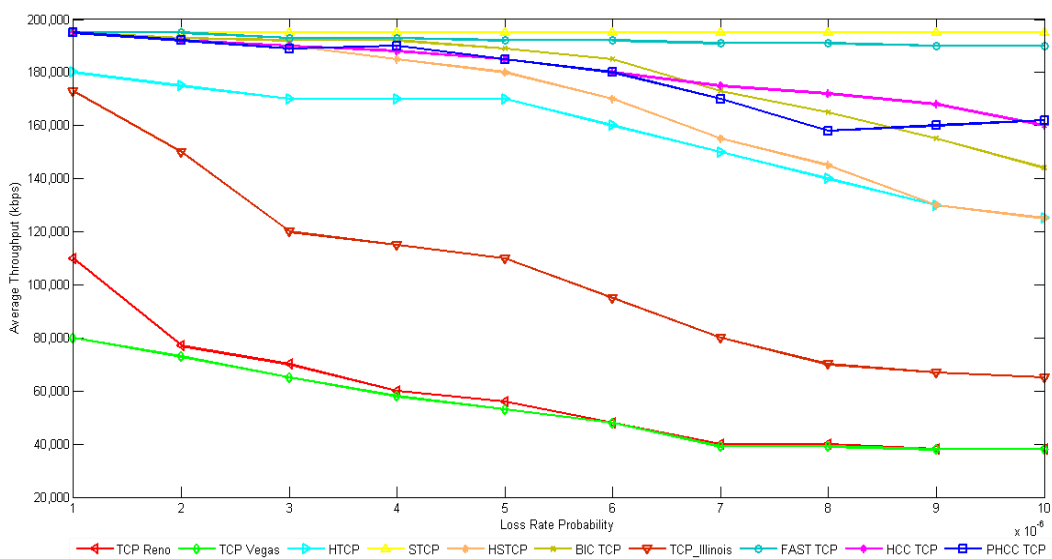

**FIGURE 10** Comparison of the packet loss vs throughput with buffer size (2000 packets).

The average queuing size versus buffer size is shown in Fig. 11. It is seen that the average queuing size of the protocols based on packet loss, such as TCP Reno, STCP, STCP, BIC TCP, and HTCP increments quickly with the buffer size growth. Therefore, the delay-based protocols utilizing the queuing delay as the primary indicator for congestion control, incur less overloads to routers compared to the loss-based protocols, particularly for the larger buffer size. But some delay-based or hybrid methods based on the queuing delay with fixed of delay time that may be another reason other than the queue is formed, not optimized for use of available buffer size. Compared with other algorithms, the proposed algorithm is optimal, because of both indicators delay and packet loss as the ratio of the probability to each other and the past are used. Fig. 11 shows the protocols average queuing size for the different buffer sizes.

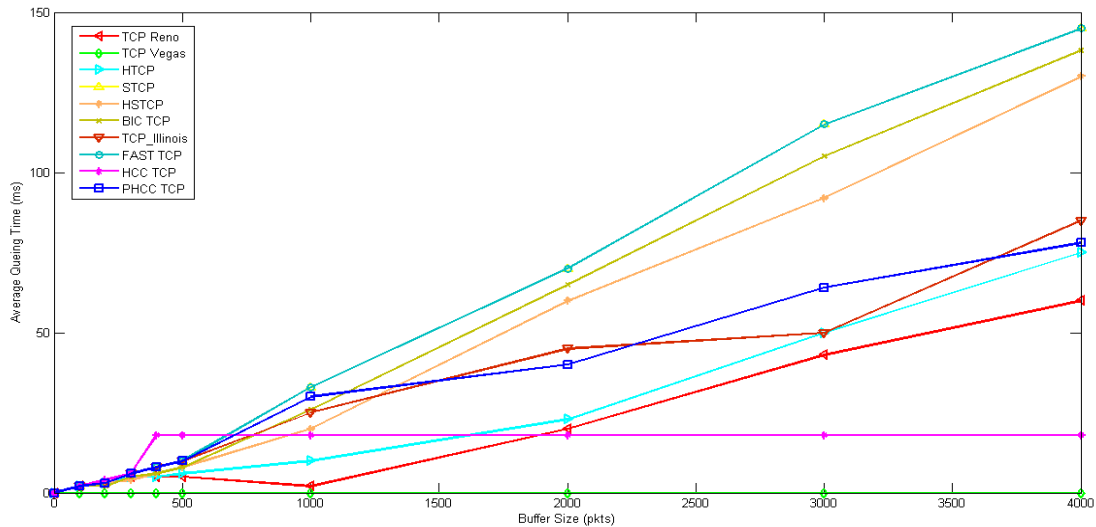

**FIGURE 11** Average queuing size versus buffer size.

To evaluate the efficiency and the use of available bandwidth by algorithm, is tested with three homogeneous RTT (120 ms), buffer size: 600 packets on each of the algorithms we've done. Average throughput obtained for calculating the utilization been used the results in Table 1 is shown. According to the results, the utilization of the proposed algorithm is 94% which is good. Although this algorithm in this regard, not above all, but in critical condition has better performance.

**TABLE 1** Comparison of the all method in terms of throughput and utilization with buffer size (600 packets).

|  | Average throughput (kbps) | Utilization |
|---|---|---|
| STCP | 195000 | 98% |
| BIC | 195000 | 98% |
| HSTCP | 188000 | 94% |
| PHCC | 187000 | 94% |
| HCC | 186000 | 93% |
| TCP-Illinois | 177000 | 89% |
| FAST | 166000 | 83% |
| HTCP | 152000 | 76% |

**Fairness:** As the first experiment to evaluate the fairness of the proposed algorithm, we use three homogeneous data flow of the proposed protocol with a buffer size of 200 packets, that at three different times to send the package. The flow 1 at 0 seconds, the flow 2 at 10 seconds and the flow 3 at 20 seconds, uses of bottleneck link to send packets. Due to the lack of competition, the first flow was in possession of all the available bandwidth shared connection. With the arrival of the second flow, available bandwidth is divided between them and also arrival third flow, fair competition between them was formed. The results shown in Figure 12(a). In Figure 12(b), the fairness of the proposed method compared to the normal TCP and high-speed TCP protocols is shown. In Continues for evaluate the fairness performance of PHCC TCP, we consider three homogeneous RTT with 120 ms and the Jain's fairness index (FI) is used to quantitatively evaluate the fairness performance of the protocols [7 and 10]. All sources start sending data at 0 s and simulation duration is set to 300s.

$$FI = \left( \frac{\left( \sum_{i=1}^{n} \overline{X}_i \right)^2}{n * \left( \sum_{i=1}^{n} \left( \overline{X}_i \right)^2 \right)} \right) \quad (11)$$

in which $n$ shows the number of the concurrent flows and $\overline{x}_i$ represents the average throughput of the flow $i$. The *FI* value is constantly no more than 1. The higher *FI* value indicates the better fairness behavior, and by the value equal to 1, the flows competing in a network obtain definitely correspondent throughput. As shown in Fig.13, PHCC TCP achieve best fairness and good utilization as other high-speed protocols in both scenarios. As observed in Fig. 13 (a), (b), all the other protocols have different behavior for buffer size of 600 and 2000 pkts. It is found that by the buffer size of 600 pkts, the fairness of FAST TCP, STCP, and TCP-Illinois degrades considerable, moreover, the utilization of HTCP and FAST TCP is low. On the other hand, by increasing the buffer size to 2000 pkts, the fairness of HTCP, STCP, FAST TCP and TCP-Illinois is considerably reduced. Though, PHCC TCP not only reaches fair sharing of link resources amongst the 3 utilization, but also keeps good bandwidth utilization.

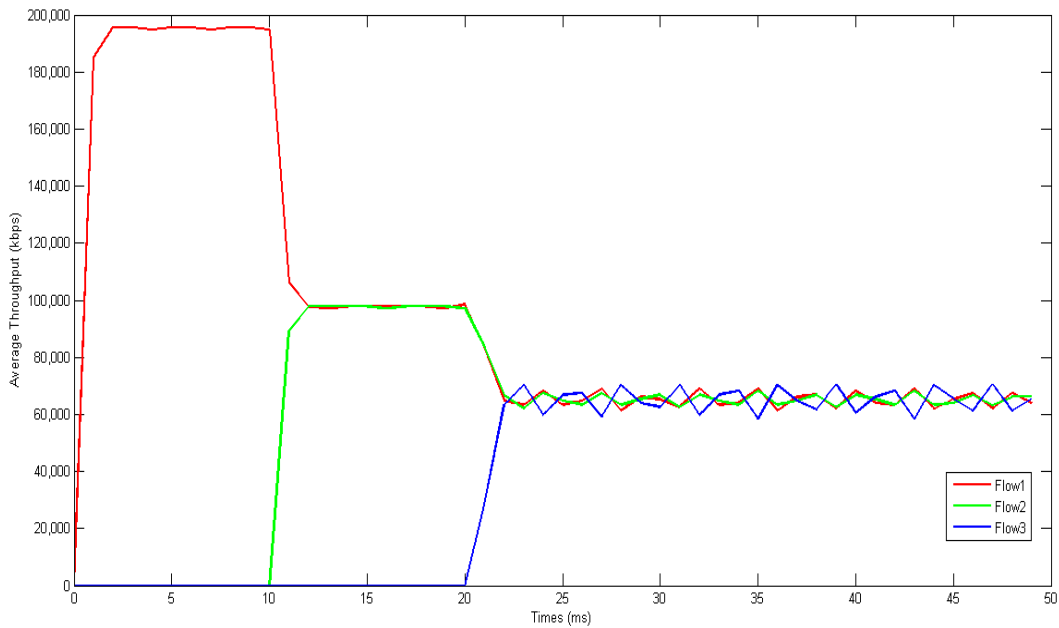

(a) 3 PHCC TCP flows.

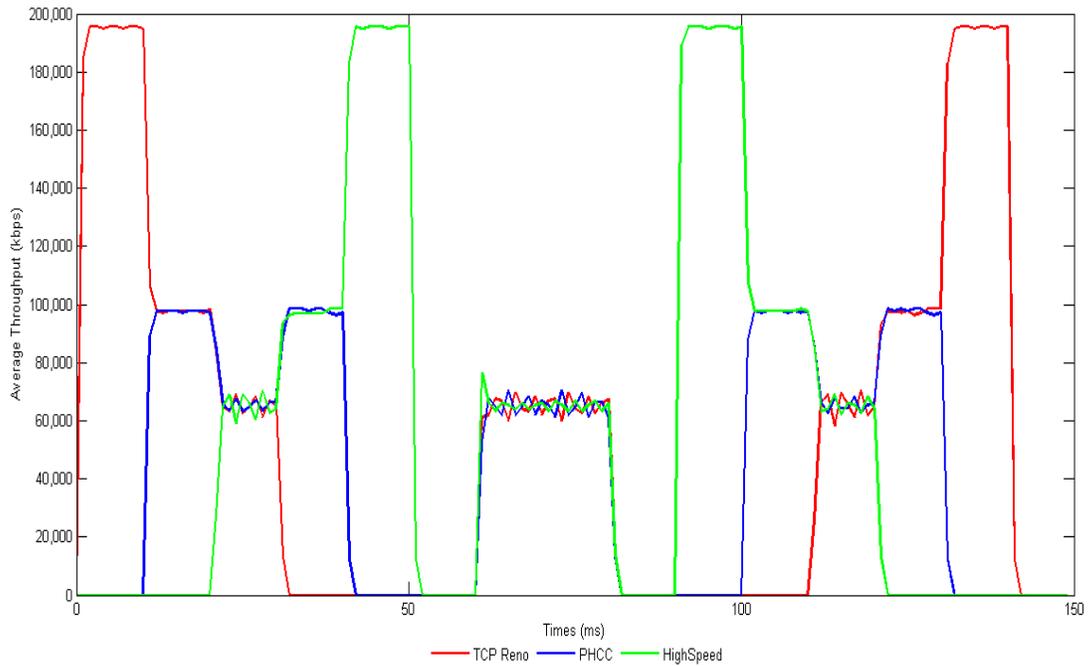

(b) PHCC, normal TCP and High-speed TCP Flows.

**FIGURE 12** Comparison of the PHCC, normal TCP and High-speed TCP flows in terms of dynamic throughput rates with buffer size (200 packets).

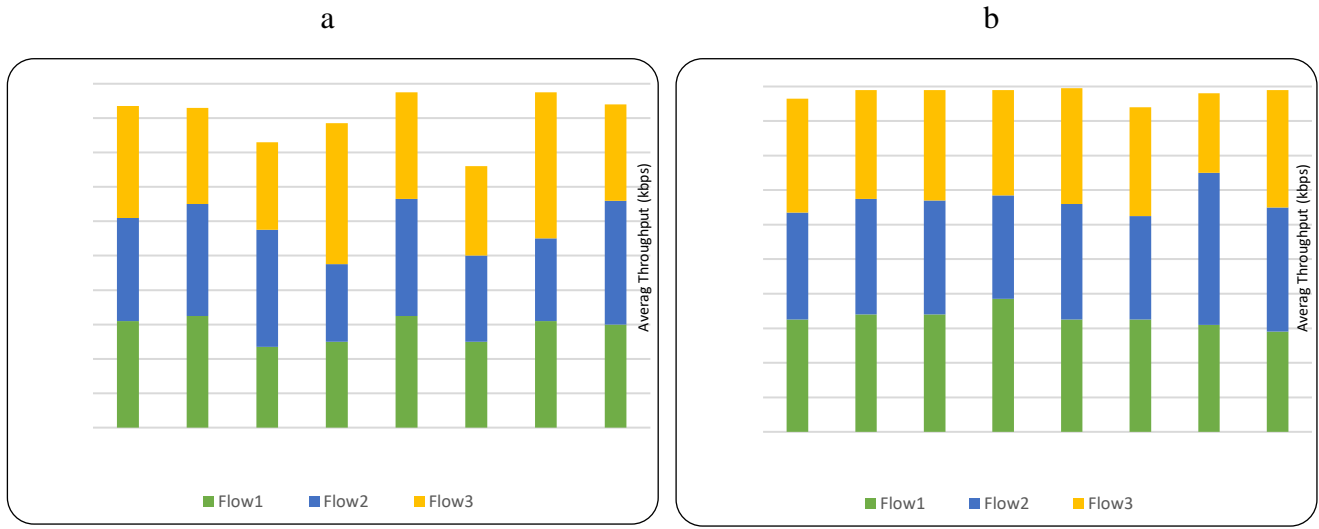

**FIGURE 13** Comparison of the 3 homogeneous RTT (120 ms) users in terms of throughput (a) Buffer size (600 packets) (b) Buffer size (2000 packets).

Fairness index values for three flow with the buffer size of 600 and 2000 pkts are calculated that results shown in Table 2. We can see that PHCC TCP is able to establish a good fairness for the three flow, the proof of this claim is the value of 1 for FI.

**TABLE 2** Comparison of the 3 homogeneous RTT (120 ms and result of FI in terms of throughput (a) Buffer size (2000 packets) (b) Buffer size (600 packets).

a

| Protocols | Average throughput (kbps) | | | Fairness Index |
|---|---|---|---|---|
| | Flow 1 | Flow 2 | Flow 3 | |
| HSTCP | 58000 | 72000 | 68000 | 0.99 |
| STCP | 62000 | 88000 | 46000 | 0.93 |
| HTCP | 65000 | 60000 | 63000 | 1 |
| BIC | 65000 | 67000 | 67000 | 1 |
| TCP-Illinois | 77000 | 60000 | 61000 | 0.99 |
| FAST | 68000 | 66000 | 64000 | 1 |
| HCC | 68000 | 67000 | 63000 | 1 |
| PHCC | 65000 | 62000 | 66000 | 1 |

b

| Protocols | Average throughput (kbps) | | | Fairness Index |
|---|---|---|---|---|
| | Flow 1 | Flow 2 | Flow 3 | |
| HSTCP | 60000 | 72000 | 56000 | 0.99 |
| STCP | 62000 | 48000 | 85000 | 0.99 |
| HTCP | 50000 | 50000 | 52000 | 1 |
| BIC | 65000 | 68000 | 62000 | 1 |
| TCP-Illinois | 50000 | 45000 | 82000 | 0.93 |
| FAST | 47000 | 68000 | 51000 | 0.97 |
| HCC | 65000 | 65000 | 56000 | 1 |
| PHCC | 62000 | 60000 | 65000 | 1 |

**TCP-friendliness:** In this subsection, we evaluate the TCP-friendliness performance, the proposed algorithm with other high-speed algorithms, compete with normal TCP Reno protocol were evaluated. We implement the simulations with two sources of the TCP Reno and two sources of the TCP variants with high speed in a homogeneous RTT setup. Figure 14 represents the average throughput of the 4 flows with various buffer sizes with the separated flow of a protocol with a definite color range. Based on the findings, it is observed that for both two size of buffers, the loss-based protocols such as STCP, HSTCP, BIC TCP, and HTCP always act unfairly and considerably decrease the average throughput of the TCP Reno flows. TCP-Illinois acts well compared to the loss-based protocols. Nevertheless, the TCP-Illinois flows reach lower throughput compared to the TCP Reno flows when the buffer size grows, HCC TCP and FAST TCP also have performance similar TCP-Illinois. In fig. 14, it is observed that with growth of buffer size, PHCC TCP have a more friendly behavior with TCP Reno flows. Based on the simulation data, it is demonstrated that PHCC TCP does not always overwhelm the concomitant TCP Reno flows and reaches superior TCP-friendliness behavior compared to the all high-speed and loss-based TCP variants.

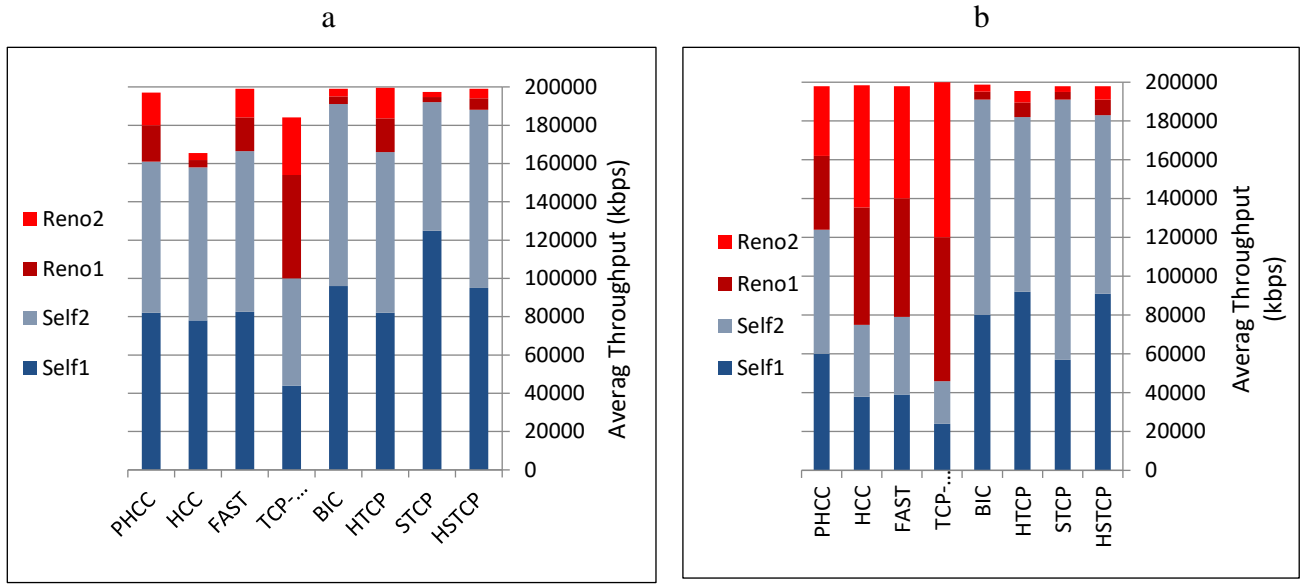

**FIGURE 14** Comparison of the TCP Reno, and high-speed TCP in terms of throughput (a) Buffer size (600 packets) (b) Buffer size (2000 packets).

## 5 | Conclusion

This paper presents an effective congestion control approach called PHCC, which uses simultaneous latency and packet-based strategies to increase data transmission performance in high-speed networks. The proposed method uses the concept of a probabilistic function and the Bayesian theorem to estimate the appropriate size of a congestion window to make the most of the available bandwidth. Because it is difficult to analyze the performance of network traffic, due to the lack of access to the values of some indicators, such as buffer size, router status and behavior of other data streams, etc. or due to lack of knowledge of the factors affecting such failures And events that may occur in the network, some of which are very complex and unpredictable, used in performance analysis. Therefore, the method used in this paper is based on probability theory, which uses the average value of the congestion window to estimate the size of the congestion window, which is calculated simultaneously based on the estimated delay and probability of loss. This feature of PHCC distinguishes it from other previously proposed algorithms and makes better decisions in acute situations. The simulation results showed that the performance of the proposed method was much more effective in throughput and TCP-friendliness criteria.

## Conflict of Interest

None.

## DATA Availability Statement

The data of this paper is the result of simulation and all the data are presented in the form of graphs inside the paper. There is no private data in this article.

## Funding

None